\documentclass[aps,pre,showpacs,amsfonts,superscriptaddress,byrevtex]{revtex4}

\usepackage{graphicx}
\usepackage{feynmf}
\begin{document}
\title{
Large scale emergent properties of an autocatalytic
reaction-diffusion model subject to noise}
\author{David Hochberg}
\email{hochberg@laeff.esa.es}
\affiliation{Centro de
Astrobiolog\'{\i}a (CSIC-INTA), Ctra. Ajalvir Km. 4, 28850
Torrej\'{o}n de Ardoz, Madrid, Spain}
\author{Felipe Lesmes}
\email{lesmeszf@inta.es} \affiliation{Centro de Astrobiolog\'{\i}a
(CSIC-INTA), Ctra. Ajalvir Km. 4, 28850 Torrej\'{o}n de Ardoz,
Madrid, Spain}
\author{Federico Mor\'{a}n}
\email{fmoran@solea.quim.ucm.es}
\affiliation{Centro de
Astrobiolog\'{\i}a (CSIC-INTA), Ctra. Ajalvir Km. 4, 28850
Torrej\'{o}n de Ardoz, Madrid, Spain} \affiliation{Departamento de
Bioqu\'{\i}mica y Biolog\'{\i}a
Molecular, Facultad de Ciencias Qu\'{\i}micas \\
Universidad Complutense de Madrid, Spain}
\author{Juan P\'{e}rez-Mercader}
\email{mercader@laeff.esa.es} \homepage[ ]{http://www.cab.inta.es}
\affiliation{Centro de Astrobiolog\'{\i}a (CSIC-INTA), Ctra.
Ajalvir Km. 4, 28850 Torrej\'{o}n de Ardoz, Madrid, Spain}

\date{\today}
\begin{abstract}
The non-equilibrium dynamic fluctuations of a stochastic version
of the Gray-Scott (GS) model are studied analytically in leading
order in perturbation theory by means of the dynamic
renormalization group. There is an attracting stable fixed point
at one-loop order, and the asymptotic scaling of the correlation
functions is predicted for both spatial and temporally correlated
noise sources. New effective three-body reaction terms, not
present in the original GS model, are induced by the combined
interplay of the fluctuations and nonlinearities.
\end{abstract}


\maketitle

\section{\label{sec:intro}Introduction}

The problem of pattern formation is a multi-disciplinary challenge
of great technological and scientific interest, and has been
studied extensively in recent years by physicists, chemists,
biologists, materials scientists and others \cite{Walgraef}. Much
of the work to date has been been devoted largely to general
aspects of pattern-forming instabilities, pattern selection, waves
and fronts, and direct numerical simulation of the idealized
\emph{deterministic} equations modelling the phenomena under study
\cite{CrossHohenberg}. In this approach to modelling, it is
reasonable to assume a continuum and \emph{coarse-grained}
description for the dynamic variables (degrees of freedom) and the
model equations are frequently, though not always, of the
reaction-diffusion type. Thus, one begins by writing down some
specific deterministic model equations. This may be regarded as
the phenomenological approach to the problem. However, a first
principles derivation of the dynamical equations is obtained by
removing the fast or short wavelength degrees of freedom from the
microscopic description, and this process leads to unavoidable
noise terms, representing the effects of internal fluctuations,
which result from this small-scale elimination step
\cite{Lee&Cardy}. Additionally, in the the study of complex
phenomena, there is typically no precise knowledge of many of the
microscopic details, nor of the initial or boundary conditions
needed to provide a \emph{complete} description of the problem.
The dynamics may also evolve in a medium (such as a background
fluid) which provides external perturbations, environmental noise
and unpredictable disturbances. Nevertheless, one is interested in
an explicit understanding of the system at long wavelengths. For
all these reasons, it is natural to consider the influence of
random noise on otherwise deterministic models and to study
stochastic reaction-diffusion equations.

One of the simplest models of biochemical relevance leading to
spatial and temporal patterns when diffusion is included is that
due to Gray and Scott \cite{Gray83}. Numerical simulations of
deterministic systems have revealed a surprisingly large set of
hitherto unknown complex and irregular patterns \cite{Pearson}.
Because of the above considerations, and in regards to these
recent findings, three questions that immediately come to mind are
(a) how do the fluctuations affect the stability of an established
pattern? and (b) what are the emergent properties, due to
fluctuation effects, of such a system at long wavelengths? and (c)
how do the deterministic and stochastic effects compete? The
purpose of this paper is to formulate carefully an analytic answer
to the second question posed in (b). The latter will be carried
out with the help of the dynamic renormalization group (RG). The
questions raised in (a) and (c), of how noise influences pattern
selection will be investigated numerically and will be presented
elsewhere. Striking numerical evidence for noise controlled
pattern self-replication is discussed in \cite{LHMP2}, where the
replication rate is maximal for an optimal but small noise
intensity.

We therefore consider a stochastic version of the Gray-Scott model
\cite{Gray83}, defined by the following system of stochastic
partial differential equations:
\begin{eqnarray}
\label{EOM} \frac{\partial}{\partial t} V &=& \lambda U V^2 - \mu
V + D_v \nabla^2
 V
+ \eta_v(\mathbf{x},t) \nonumber\\
\frac{\partial}{\partial t} U &=& u_0 - \lambda U V^2 - \nu U +
D_u
 \nabla^2 U
+ \eta_u(\mathbf{x},t).
\end{eqnarray}
In the \emph{absence} of noise, (\ref{EOM}) coincides with the
Gray-Scott model \cite{Gray83}, which is a variant of the
autocatalytic Selkov model of glycolysis, corresponding to the
following chemical reactions:
\begin{eqnarray}\label{reaction}
U + 2 V &\stackrel{\lambda}{\to}& 3 V, \nonumber\\
V &\stackrel{\mu}{\to}& P, \nonumber\\
U &\stackrel{\nu}{\to}& Q, \nonumber\\
  &\stackrel{u_0}{\to}& U.
\end{eqnarray}
$V(\mathbf{x},t)$ and $U(\mathbf{x},t)$ represent the
concentrations of the chemical species $U$ and $V$, and are
functions of $d$-dimensional space $\mathbf{x}$ and time $t$.
$\lambda$ is the reaction rate, $P$ and $Q$ are inert products,
$\mu$ is the decay rate of $V$ and $\nu$ is the decay rate of $U$
and $u_0$ is the constant feed rate. In \cite{Pearson}, the
reaction rate parameter was simply set equal to unity: $\lambda =
1$. Here we retain it as a free parameter, as we will see that it
undergoes a nontrivial renormalization.  A non-equilibrium
constraint is represented by a feed term for $U$. The rate at
which $U$ is supplied is positive if the concentration of $U$
drops below an equilibrium value and negative if it exceeds it.
The equilibrium $U$ concentration is $u_0/\nu$, where $u_0$ is the
feed rate constant. The chemical species $U$ and $V$ can diffuse
with independent diffusion constants $D_u$ and $D_v$. All the
model parameters are positive.

Any real chemical system is subjected to random fluctuations. We
can include such effects in the GS model by means of noise terms,
that can in principle be additive or multiplicative. In this work,
we have chosen to investigate the influence of additive noise
alone as a initial approach to the full stochastic problem in
which both types of fluctuations can be simultaneously present.
This is intended therefore as a first step towards incorporating
fluctuation effects. This choice is influenced in part by the
technical aspects of the dynamical renormalization group. From
this perspective, additive noise, which enters the dynamics
linearly, is technically easier to treat analytically in RG
calculations. Accordingly, here we have opted to use independent
additive colored noises, $\eta_v(\mathbf{x},t)$ and
$\eta_u(\mathbf{x},t)$. First, spatially-correlated noise, and
then temporally-correlated noise are considered. A few cautionary
remarks regarding our incorporation of noise are in order. We
emphasize that here we are adopting a simple phenomenological
approach, wherein noise is added to the idealized deterministic
equations. But, stochastic equations can also be derived from
first principles. We refer to the methods that take a classical
master equation, purporting to encode the precise microchemistry,
to a continuum field theory \cite{Doi,Peliti,Grassberger}. This
"second-quantized" formalism requires detailed knowledge of the
microscopic master equation, and the end result of this method
yields the stochastic PDE for the coarse-grained degrees of
freedom and the (internal) noise correlations. A technical
discrepancy comes up in applying phenomenological
reaction-diffusion equations with real additive noise to
pair-reaction kinetics (i.e. $V + V \to 0$), since in this
particular case of particle annihilation, the SPDE derived from
the fundamental microscopic master equation is \emph{complex} and
contains \emph{imaginary} noise \cite{Howard}. However, for
standard Gribov processes (particle clustering reactions), the
first-principles method does lead to a real SPDE with real
additive noise. Thus, the ad-hoc method is adequate for handling
these cases.  For external or environmental noise, the
phenomenological strategy adopted here is also adequate since
external fluctuations are typically specified only at the
coarse-grained level.

The remainder of the paper is organized as follows. In the next
section, we specify the noise properties directly in Fourier space
and derive the scaling laws that the model parameters in
(\ref{EOM}) obey when the stochastic equations are scale invariant
(as they are in the neighborhood of any RG fixed point) and the
general scaling form of the correlation functions of the
composition fields is deduced in Subsection \ref{sec:corr}. This
information will be needed when we apply the renormalization group
in Section \ref{sec:rg}. The fixed points and associated exponents
are solved for and the RG flow is represented in a reduced two
dimensional parameter space. The global flow patterns clearly
illustrate the $d$-dependence of the flow, the role of the
associated critical dimension, as well as the phenomena of
crossover. In Section \ref{sec:new3}, we discuss the emergence of
effective three-body reaction terms not present in the original GS
model that arise as a direct consequence of fluctuations and
non-linearity. Modifications and changes in the RG results due to
the presence of temporally correlated noise are very briefly
discussed in Section \ref{sec:tempcor}. Conclusions and discussion
are presented in Section \ref{sec:disc}.

\section{\label{sec:naive} Naive Scaling Properties}

We can deduce the naive scaling laws that must hold if the
stochastic equations (\ref{EOM}) are to be \emph{form-invariant}
under a basic re-scaling of both space and time. To do so, we
specify the properties of the noise. In Fourier space, the noise
correlations are given by
\begin{eqnarray}\label{noisecorr}
\left\langle \eta_v(\mathbf{k},\omega) \eta_v(\mathbf{k'},\omega')
 \right\rangle &=&
2 (2 \pi)^{d+1} A_v k^{-y_v} \delta^d(\mathbf{k}+\mathbf{k'})
\delta(\omega+\omega')\nonumber\\
\left\langle \eta_u(\mathbf{k},\omega) \eta_u(\mathbf{k'},\omega')
 \right\rangle &=&
2 (2 \pi)^{d+1} A_u k^{-y_u} \delta^d(\mathbf{k}+\mathbf{k'})
\delta(\omega+\omega')\nonumber \\
\left\langle \eta_v(\mathbf{k},\omega) \eta_u(\mathbf{k'},\omega')
 \right\rangle &=& 0.
\end{eqnarray}
We assume all cumulants except the above to vanish, thus the
noises are individually Gaussian, centered around zero and
mutually uncorrelated. The independent noise amplitudes $A_v,A_u
> 0$ are positive definite and the noise exponents $y_v$ and $y_u$
are real free parameters.

Space, time and the concentrations $U$, $V$, are re-scaled by the
following transformations:
\begin{eqnarray}
\mathbf{x} &\to& s^{-1} \mathbf{x} \nonumber \\
t &\to& s^{-z} t \nonumber \\
V &\to& s^{-\chi_v} V \nonumber \\
U &\to& s^{-\chi_u} U.
\end{eqnarray}
Here, $s>1$ is a convenient scale factor and $z$ is the so-called
dynamic exponent. We initially allow for each concentration field
to respond independently under the re-scaling; this is reflected
by the two a-priori independent roughness exponents $\chi_v$ and
$\chi_u$, respectively. However, the fact that the reaction rate
$\lambda$ appears simultaneously in both the $U$ and $V$ equation
requires that $\chi_u = \chi_v \equiv \chi$, as is easy to check.
There is just a single roughness exponent to deal with.

Subjecting the equations (\ref{EOM}) to this transformation, we
find the model parameters scale naively (classically) in the
following way,
\begin{eqnarray}\label{naive}
\nu &\to& s^z \nu \nonumber \\
\mu &\to& s^z \mu \nonumber \\
D_v &\to& s^{z-2} D_v \nonumber \\
D_u &\to& s^{z-2} D_u \nonumber \\
u_0 &\to& s^{z-\chi} u_0 \nonumber \\
A_v &\to& s^{y_v-d+z-2\chi} A_v \nonumber \\
A_u &\to& s^{y_u-d+z-2\chi} A_u \nonumber \\
\lambda &\to& s^{2 \chi + z} \lambda_u. \nonumber \\
\end{eqnarray}

\subsection{\label{sec:corr} Correlation functions}

Both the dynamic and roughness exponents govern the scaling form
of the correlation functions of the concentration fields. To see
this, consider the effect of a scale transformation on the
$V$-field which we write more explicitly as
\begin{equation}
V(s\mathbf{x},s^z t) = s^{\chi} V(\mathbf{x},t),
\end{equation}
which holds in the scaling regime (when the system is near one of
its RG fixed points). Then the correlation function will scale as
\begin{eqnarray}\label{asymp1}
\langle V(\mathbf{x},t)\,V(0,0)\rangle &=& s^{-2\chi}\,\langle V(s
\mathbf{x},s^z t)\,V(0,0)\rangle \nonumber \\
&=& |\mathbf{x}|^{2\chi}\, \Phi \left( \frac{t}{|\mathbf{x}|^z}
\right),
\end{eqnarray}
where the scaling function $\Phi$ obeys the following asymptotic
limits \cite{Frey-Tauber}
\begin{equation}\label{asymp2}
\lim_{u \to 0} \Phi(u) = {\rm const.} \qquad \lim_{u \to \infty}
\Phi(u) = {\rm const.} \, \times u^{2\chi/z}.
\end{equation}
Thus, knowledge of both scaling exponents $(z,\chi)$ is required
in order to predict the long-wavelength and long-time correlations
of the fields. Similar considerations apply to the $\langle UU
\rangle$ and cross-correlation $\langle VU \rangle$. In fact, the
scaling of these latter two correlations will obey a relation
similar to (\ref{asymp1}), except for a possibly different scaling
function. Nevertheless, the limits in (\ref{asymp2}) are valid for
all scaling functions, although the constants appearing there can
be distinct. The tool best suited for the calculation of the
required exponents is provided by the renormalization group, to
which we next turn.

\section{\label{sec:rg} Renormalization Group (RG) Analysis}

Since we are interested in the emergent scaling properties of the
model, we focus on the hydrodynamic limit, that is, the long-time
$t \to \infty$ and long-distance $ \mathbf{x} \to \infty$ limits.
The dynamical RG \cite{Hohenberg77,Ma76} is a powerful tool for
computing and analyzing the asymptotic properties of
out-of-equilibrium stochastic systems.

In order to calculate the scaling exponents in the asymptotic
limit, the linear and non-linear parts of the equations of motion
can be re-organized so as to allow a perturbative calculation in
powers of the non-linear parameter $\lambda$. After Fourier
transforming, the coupled stochastic equations of motion
(\ref{EOM}) can be rewritten as
\begin{eqnarray}
\label{Formal_solution} \lefteqn{V(\mathbf{k},\omega) =
  {G_v}_0(\mathbf{k},\omega) \eta_v(\mathbf{k},\omega) + } \nonumber \\
  & & {G_v}_0(\mathbf{k},\omega) \lambda
      \int \frac{d^d \mathbf{k_1}}{{(2\pi)}^d} \frac{d \omega_1}{2\pi}
           \frac{d^d \mathbf{k_2}}{{(2\pi)}^d} \frac{d \omega_2}{2\pi}
        V(\mathbf{k_1},\omega_1) V(\mathbf{k_2},\omega_2)
        U(\mathbf{k}-\mathbf{k_1}-\mathbf{k_2},\omega-\omega_1-\omega_2) \nonumber \\
\lefteqn{U(\mathbf{k},\omega) =
  {G_u}_0(\mathbf{k},\omega) \eta_u(\mathbf{k},\omega) +
  {G_u}_0(\mathbf{k},\omega) {(2\pi)}^{d+1} \delta^d(\mathbf{k}) \delta(\omega) u_0 - }\\
  & & {G_u}_0(\mathbf{k},\omega) \lambda
      \int \frac{d^d \mathbf{k_1}}{{(2\pi)}^d} \frac{d \omega_1}{2\pi}
           \frac{d^d \mathbf{k_2}}{{(2\pi)}^d} \frac{d \omega_2}{2\pi}
        V(\mathbf{k_1},\omega_1) V(\mathbf{k_2},\omega_2)
        U(\mathbf{k}-\mathbf{k_1}-\mathbf{k_2},\omega-\omega_1-\omega_2), \nonumber
\end{eqnarray}
where the bare propagators (or, response functions) ${G_v}_0$ and
${G_u}_0$ are defined by
\begin{eqnarray}\label{Bare_propagator}
{G_v}_0(\mathbf{k},\omega) &=&
  \frac{1}{\mu + D_v  k^2 - i\omega } \nonumber \\
{G_u}_0(\mathbf{k},\omega) &=&
  \frac{1}{\nu + D_u k^2 - i\omega }.
\end{eqnarray}
The wave-number modulus is denoted by $k = |\mathbf{k}|$. Define
the effective propagators by
\begin{eqnarray}\label{effpropV}
V(\mathbf{k},\omega) &\equiv&
{G_v}(\mathbf{k},\omega)\eta_v(\mathbf{k},\omega) \\
\label{effpropU} U(\mathbf{k},\omega) &\equiv&
{G_u}(\mathbf{k},\omega) \{ \eta_u(\mathbf{k},\omega) +
(2\pi)^{d+1} \delta^d(\mathbf{k}) \delta(\omega) u_0 \}.
\end{eqnarray}
Substituting these into the set of integral equations
(\ref{Formal_solution}) yields a set of equations that can be
solved iteratively to any order in $\lambda$. To this end, it is
best to handle the expansion via diagrams, and the above equations
(\ref{Formal_solution}) are represented diagrammatically as
follows:
\begin{fmffile}{fmfdiagrams}
\fmfcmd{%
  style_def charged_boson expr p =
   draw (wiggly p);
   fill (arrow p)
  enddef;}
\begin{eqnarray*}\label{eomgraphs}
\mbox{\parbox{24mm}{
\begin{fmfchar*}(20,15)
  \fmfleft{i1}
  \fmfright{o1}
  \fmflabel{$\eta_v$}{o1}
  \fmf{double_arrow}{i1,o1}
\end{fmfchar*}
}} \quad &=& \mbox{\parbox{24mm}{
\begin{fmfchar*}(20,15)
  \fmfleft{i1}
  \fmfright{o1}
  \fmflabel{$\eta_v$}{o1}
  \fmf{fermion}{i1,o1}
\end{fmfchar*}
}} \qquad + \quad \mbox{\parbox{37mm}{ \vspace{5mm}
\begin{fmfchar*}(30,15)
  \fmfleft{i1}
  \fmfright{o3,o2,o1}
  \fmflabel{$j+\otimes$}{o3}
  \fmflabel{$\eta_v$}{o2}
  \fmflabel{$\eta_v$}{o1}
  \fmf{fermion,tension=5}{i1,v1}
  \fmf{double_arrow}{v1,o1}
  \fmf{double_arrow}{v1,o2}
  \fmf{dbl_wiggly}{v1,o3}
  \fmf{phantom_arrow}{v1,o3}
  \fmfv{d.sh=circle,d.filled=empty,label=$1$,label.dist=0}{v1}
\end{fmfchar*}
\vspace{5mm}
}} \\
\mbox{\parbox{30mm}{
\begin{fmfchar*}(20,15)
  \fmfleft{i1}
  \fmfright{o1}
  \fmflabel{$\eta_u +\otimes$}{o1}
  \fmf{dbl_wiggly}{i1,o1}
  \fmf{phantom_arrow}{i1,o1}
\end{fmfchar*}
}} \quad &=& \mbox{\parbox{30mm}{
\begin{fmfchar*}(20,15)
  \fmfleft{i1}
  \fmfright{o1}
  \fmflabel{$\eta_u +\otimes$}{o1}
  \fmf{charged_boson}{i1,o1}
\end{fmfchar*}
}} \qquad + \quad \mbox{\parbox{37mm}{ \vspace{5mm}
\begin{fmfchar*}(30,15)
  \fmfleft{i1}
  \fmfright{o3,o2,o1}
  \fmflabel{$\eta_u +\otimes$}{o3}
  \fmflabel{$\eta_v$}{o2}
  \fmflabel{$\eta_v$}{o1}
  \fmf{charged_boson,tension=5}{i1,v1}
  \fmf{double_arrow}{v1,o1}
  \fmf{double_arrow}{v1,o2}
  \fmf{dbl_wiggly}{v1,o3}
  \fmf{phantom_arrow}{v1,o3}
  \fmfv{d.sh=circle,d.filled=empty,label=$2$,label.dist=0}{v1}
\end{fmfchar*}
\vspace{5mm} }}
\end{eqnarray*}
The directed double straight (wiggly) line symbol represents the
effective $V$-propagator ($U$-propagator) defined above: straight
and wiggly lines correspond to the $V$-sector, and $U$-sector,
respectively. The single directed lines stand for the bare
propagators (\ref{Bare_propagator}). It is understood that each
directed line carries wave number $\mathbf{k}$ and frequency
$\omega$, which is conserved at all vertices, as can be seen from
(\ref{Formal_solution}). We do not write down this dependence
explicitly on the graphs in order to avoid clutter. The non-linear
coupling terms, or vertices, are represented by the directed
four-pronged symbols, of which there are two, and these are
denoted by the encircled numerals \textcircled{1} and
\textcircled{2}, respectively. The zero-mode $u_0$ in the second
equation in (\ref{Formal_solution}) is indicated above by a
$\bigotimes$. The perturbation expansion can now be performed
efficiently without having to carry along lengthy and tedious
algebraic expressions. For further details on using graphical
methods to solve stochastic differential equations, see the
Appendices of \cite{Barabasi}. The calculation of the propagator
then follows from a graphical iteration of these expressions,
amputating one noise factor from the legs, followed by an
averaging over the remaining noise factors. The results at
\emph{one-loop} level are indicated as follows:
\begin{eqnarray*}\label{propgraphs}
\mbox{\parbox{20mm}{
\begin{fmfchar*}(20,15)
  \fmfleft{i1}
  \fmfright{o1}
  \fmf{double_arrow}{i1,o1}
\end{fmfchar*}
}} \quad &=& \mbox{\parbox{20mm}{
\begin{fmfchar*}(20,15)
  \fmfleft{i1}
  \fmfright{o1}
  \fmf{fermion}{i1,o1}
\end{fmfchar*}
}}
\quad + \quad \mbox{2 loops} \\
\mbox{\parbox{20mm}{
\begin{fmfchar*}(20,15)
  \fmfleft{i1}
  \fmfright{o1}
  \fmf{dbl_wiggly}{i1,o1}
  \fmf{phantom_arrow}{i1,o1}
\end{fmfchar*}
}} \quad &=& \mbox{\parbox{20mm}{
\begin{fmfchar*}(20,15)
  \fmfleft{i1}
  \fmfright{o1}
  \fmf{boson}{i1,o1}
  \fmf{phantom_arrow}{i1,o1}
\end{fmfchar*}
}} \quad + \quad \mbox{\parbox{30mm}{ \vspace{5mm}
\begin{fmfchar*}(30,15)
  \fmfleft{i1,i2}
  \fmfright{o1,o2}
  \fmftop{t1}
  \fmf{charged_boson}{i1,v1}
  \fmf{charged_boson}{v1,o1}
  \fmf{fermion,left,tension=0}{v1,t1}
  \fmf{fermion,right,tension=0}{v1,t1}
  \fmfv{d.sh=circle,d.filled=empty,label=$2$,label.dist=0}{v1}
  \fmfv{d.sh=circle,d.filled=empty,label=$\eta\eta$,label.dist=0}{t1}
\end{fmfchar*}
\vspace{5mm} }}
\quad + \quad \mbox{2 loops} \\
\end{eqnarray*}
Note that only the $U$ propagator receives corrections at
one-loop, whereas the $V$ propagator is unchanged to this order.
Furthermore, the $v$-noise is what "drives" the renormalization of
the $U$-propagator. The loop expansion for the vertices is
indicated below. One takes the vertices and iterates, to a desired
order in the coupling $\lambda$,  by replacing directed double
lines using the above equations of motion. Then, three noise
factors (two $\eta_v$'s and one $\eta_u$) are amputated and the
resultant graphs are averaged over the remaining noises. A
combinatorial factor of $4$ appears after counting all possible
noise contractions. We thus obtain,
\begin{eqnarray*}\label{vertexgraphs}
\parbox{30mm}{
\begin{fmfchar*}(30,15)
  \fmfleft{i1}
  \fmfright{o3,o2,o1}
  \fmf{fermion,tension=5}{i1,v1}
  \fmf{double_arrow}{v1,o1}
  \fmf{double_arrow}{v1,o2}
  \fmf{dbl_wiggly}{v1,o3}
  \fmf{phantom_arrow}{v1,o3}
  \fmfv{d.sh=circle,d.filled=empty,label=$1$,label.dist=0}{v1}
\end{fmfchar*}
} &=&
\parbox{30mm}{
\begin{fmfchar*}(30,15)
  \fmfleft{i1}
  \fmfright{o3,o2,o1}
  \fmf{fermion,tension=5}{i1,v1}
  \fmf{fermion}{v1,o1}
  \fmf{fermion}{v1,o2}
  \fmf{charged_boson}{v1,o3}
  \fmfv{d.sh=circle,d.filled=empty,label=$1$,label.dist=0}{v1}
\end{fmfchar*}
} \quad + 4
\parbox{40mm}{
\vspace{2mm}
\begin{fmfchar*}(40,20)
  \fmfleft{o1,i1}
  \fmfright{o3,o2}
  \fmftop{f1,n1,f2}
  \fmf{fermion,tension=3}{i1,v1}
  \fmf{fermion,tension=3}{v1,o1}
  \fmf{charged_boson,right=0.5,tension=.3}{v1,v2}
  \fmf{fermion,left=0.5,tension=.3}{v1,n1}
  \fmf{fermion,right=0.5,tension=.3}{v2,n1}
  \fmf{fermion,tension=3}{v2,o2}
  \fmf{charged_boson,tension=3}{v2,o3}
  \fmfdot{v1,v2}
  \fmfv{d.sh=circle,d.filled=empty,d.size=100,label=$1$,label.dist=0}{v1}
  \fmfv{d.sh=circle,d.filled=empty,d.size=100,label=$2$,label.dist=0}{v2}
  \fmfv{d.sh=circle,d.filled=empty,label=$\eta\eta$,label.dist=0}{n1}
\end{fmfchar*}
\vspace{2mm} } \quad + \quad \mbox{2 loops}
\\
\parbox{30mm}{
\begin{fmfchar*}(30,15)
  \fmfleft{i1}
  \fmfright{o3,o2,o1}
  \fmf{charged_boson,tension=5}{i1,v1}
  \fmf{double_arrow}{v1,o1}
  \fmf{double_arrow}{v1,o2}
  \fmf{dbl_wiggly}{v1,o3}
  \fmf{phantom_arrow}{v1,o3}
  \fmfv{d.sh=circle,d.filled=empty,label=$2$,label.dist=0}{v1}
\end{fmfchar*}
} &=&
\parbox{30mm}{
\begin{fmfchar*}(30,15)
  \fmfleft{i1}
  \fmfright{o3,o2,o1}
  \fmf{charged_boson,tension=5}{i1,v1}
  \fmf{fermion}{v1,o1}
  \fmf{fermion}{v1,o2}
  \fmf{charged_boson}{v1,o3}
  \fmfv{d.sh=circle,d.filled=empty,label=$2$,label.dist=0}{v1}
\end{fmfchar*}
} \quad + 4
\parbox{40mm}{
\vspace{2mm}
\begin{fmfchar*}(40,20)
  \fmfleft{o1,i1}
  \fmfright{o3,o2}
  \fmftop{f1,n1,f2}
  \fmf{charged_boson,tension=3}{i1,v1}
  \fmf{fermion,tension=3}{v1,o1}
  \fmf{charged_boson,right=0.5,tension=.3}{v1,v2}
  \fmf{fermion,left=0.5,tension=.3}{v1,n1}
  \fmf{fermion,right=0.5,tension=.3}{v2,n1}
  \fmf{fermion,tension=3}{v2,o2}
  \fmf{charged_boson,tension=3}{v2,o3}
  \fmfdot{v1,v2}
  \fmfv{d.sh=circle,d.filled=empty,d.size=100,label=$2$,label.dist=0}{v1}
  \fmfv{d.sh=circle,d.filled=empty,d.size=100,label=$2$,label.dist=0}{v2}
  \fmfv{d.sh=circle,d.filled=empty,label=$\eta\eta$,label.dist=0}{n1}
\end{fmfchar*}
\vspace{2mm} }
\quad + \quad \mbox{2 loops} \\
\end{eqnarray*}
The noise spectral functions receive no corrections at one loop
order.
\end{fmffile}
We point out that although $\lambda$ is employed as a formal
expansion parameter, the bona-fide and \emph{dimensionless}
perturbation parameter $g$ (see (\ref{dimensionless}) below) is a
combination of this, a noise amplitude, a decay constant, a
diffusion constant and a short distance (or ultraviolet) cutoff
needed for convergence of the integrals. Moreover, we expand in
loops rather than in powers of $g$. Physically, the loop expansion
is an expansion in powers of the noise amplitude \cite{HMPVa}. We
note that at leading order in loops, the perturbation expansion
indicates that only the $U$ field propagator and the non-linear
coupling $\lambda$ receive corrections. The noise amplitudes and
the $V$ propagator are unchanged at one-loop order. At two-loop
order, all propagators, the noise amplitudes as well as the
vertices, do receive corrections. However, there are some
important issues at leading order that need to be understood.

The dynamical RG transformation is carried out in two steps
\cite{Ma76}:
\begin{enumerate}
\item
High momenta components, in the momentum shell $\Lambda e^{-\delta
l} < |\mathbf{k}| < \Lambda$, at and below the cutoff $\Lambda$,
are integrated out. Note: $s = e^l$.
\item
A change of scale restores the cutoff to the value $\Lambda$.
\end{enumerate}

Further details of the calculation are given in Appendices A and
B. Working with infinitesimal parameter $\delta l$ gives rise to
differential equations governing the RG flow in parameter space:
\begin{eqnarray}\label{diffRG}
\frac{d\nu}{dl} &=& z \nu +
        \frac{\lambda A_v K_d
 \Lambda^{d-y_v}}{(\mu+D_v\Lambda^2)}
        \nonumber \\
\frac{d\mu}{dl} &=& z \mu \nonumber \\
\frac{dD_v}{dl} &=& (z-2) D_v \nonumber \\
\frac{dD_u}{dl} &=& (z-2) D_u \nonumber \\
\frac{du_0}{dl} &=& (z-\chi) u_0 \nonumber \\
\frac{dA_v}{dl} &=& (y_v-d+z-2\chi) A_v \nonumber \\
\frac{dA_u}{dl} &=& (y_u-d+z-2\chi) A_u \nonumber \\
\frac{d\lambda}{dl}  &=&
        \left[
        2 \chi + z -
        \frac{4 \lambda A_v K_d \Lambda^{d-y_v} }
             {\left(\mu+D_v\Lambda^2\right)
          \left(\mu+\nu+D_v\Lambda^2 + D_u\Lambda^2\right)}
    \right] \lambda .
\end{eqnarray}
$K_d = S_d/{(2\pi)}^d $ where $S_d$ is the surface area of a
$d$-dimensional sphere. As already noted in the corresponding
graphs, at one-loop order, only two out of the total of eight
model parameters run with scale, namely, the decay rate $\nu$, of
the $U$ field and the nonlinear coupling $\lambda$. Note moreover
that at this order, the corrections are driven by the $v$-noise,
but not the $u$-noise. Despite the relative simplicity of these
equations, non-trivial RG fixed points and flow already result at
this leading order perturbation.

\subsection{\label{sec:fps} Fixed Points and Dynamic Scaling}

We determine the fixed points and associated exponents implied by
the above equations (\ref{diffRG}). We can analyze the RG flow in
a reduced two-dimensional parameter space by introducing the pair
of dimensionless couplings defined by
\begin{eqnarray}\label{dimensionless}
g &=& \lambda A_v K_d \frac{\Lambda^{d - y_v -2}}{\nu D_v},\\
h &=& \big(D_v + D_u \big)\frac{\Lambda^2}{\nu},
\end{eqnarray}
for then the RG equations (\ref{diffRG}) can be written as
\begin{eqnarray}\label{diffRG2}
\dot \nu &=& \nu \left( z + g \right),\\\label{diffRG3} \dot
\lambda &=& \lambda \left( 2\chi + z - \frac{4 g}{1 + h} \right);
\end{eqnarray}
we do not write out the remaining trivial RG equations. Note we
have taken $\mu = 0$ from the outset. This choice is necessary in
order to obtain a non-trivial $g \neq 0$ fixed point solution of
(\ref{diffRG}). Here the over-dot stands for the derivative
$d/dl$. In terms of $g$ and $h$, the nontrivial part of the RG
flow is governed by the pair of equations
\begin{eqnarray}\label{newpairg}
\dot g &=& g\, \left(y_v -d+2 -g -\frac{4 g}{1 + h} \right),\\
\label{newpairh} {\dot h} &=& -\big(2 + g \big) h.
\end{eqnarray}
The fixed points can be solved for by looking for all zeroes of
the pair (\ref{newpairg},\ref{newpairh}), to be denoted as $(g^*,
h^*)$ while the associated fixed point exponents are obtained by
substituting the solutions of $\dot g = 0$ and $\dot h = 0$ into
(\ref{diffRG2},\ref{diffRG3}); solving for the zeroes of this
latter pair then yields the exponents, denoted by $(z^*,\chi^*)$.

We first search for all non-trivial fixed points. These correspond
to $g^* \neq 0$, since $g$ is proportional to the non-linear
coupling $\lambda$. There are two cases to be distinguished.

\noindent \underline{Case(a)}: $g^* = -2$ and $1 + h^* = 8/(d-y_v
-4)$. The associated exponents are $z^* = 2$ and $\chi^* =
(2-d+y_v)/2$. Note that the combination $(z^*-2\chi^*-d+y_v)
\equiv 0$ vanishes identically. Thus, if we choose $u_0 = 0$, and
$y_u = y_v$, all remaining RG equations (\ref{diffRG}) are
stationary. Note for this fixed point solution, the stochastic GS
model is in the same universality class as the linear
Edwards-Wilkinson model \cite{Barabasi}.

\noindent \underline{Case(b)}: $h^* = 0$ and $g^* = (y_v - d
+2)/5$. The exponents associated to this fixed point are $z^* =
-(y_v - d + 2)/5$ and $\chi^* = (y_v - d + 2)/2$. Again we choose
$u_0 = 0$, then all remaining RG equations are automatically
stationary except that the noise amplitudes decay to zero on
approaching this fixed point: $A_v(l) = A_v e^{l(z-2\chi - d +
y_v) } \to 0$ as $l \to \infty$ and similarly for $A_u(l)$. This
limit holds provided that $d < 12 + y_v$, and $d < 12 + y_u$ for
the limit of $A_u$.

Since the model (\ref{EOM}) is defined as having all non-negative
parameters, we must exclude the solution in case(a) as being
physically spurious. This leaves us with the nontrivial fixed
point of case(b), which is an attractive fixed point and is
indicated by the symbol \textbf{A} in the flow graph; refer to
Fig.(\ref{flow1}).

There is one trivial fixed point: this is a saddle point, as
indicated by \textbf{S} in the flow graph Fig.(\ref{flow1}). This
corresponds to $g^* = 0$ and $h^* = 0$. The exponents are found to
be $z^* = 0$ and $\chi^* = (y_v - d)/2$. This is a fixed point
solution to all equations in (\ref{diffRG}) provided $y_v = y_u$
and again upon setting $u_0 = 0$.

\subsection{\label{sec:rg_flow}RG flow}

\begin{figure}
\includegraphics[scale=0.5,angle=270]{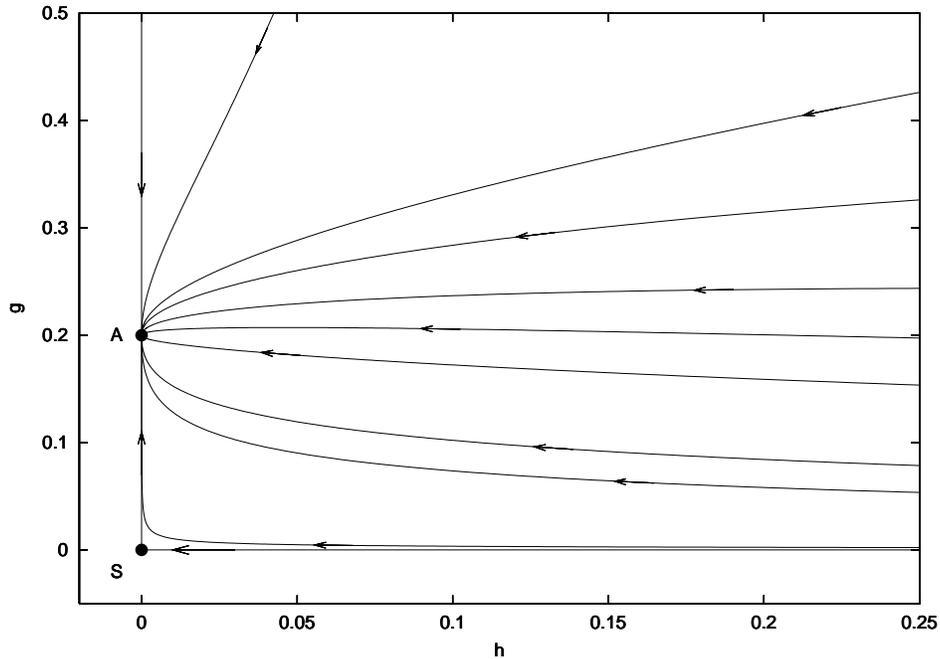}
\caption{\label{flow1} A projection of the RG flow in Eqs.
(\ref{newpairg},\ref{newpairh}). Two fixed points: \textbf{A} is
attractive, \textbf{S} is a saddle point. This is plotted for
$\epsilon > 0$. }
\end{figure}

\begin{figure}
\includegraphics[scale=0.5,angle=270]{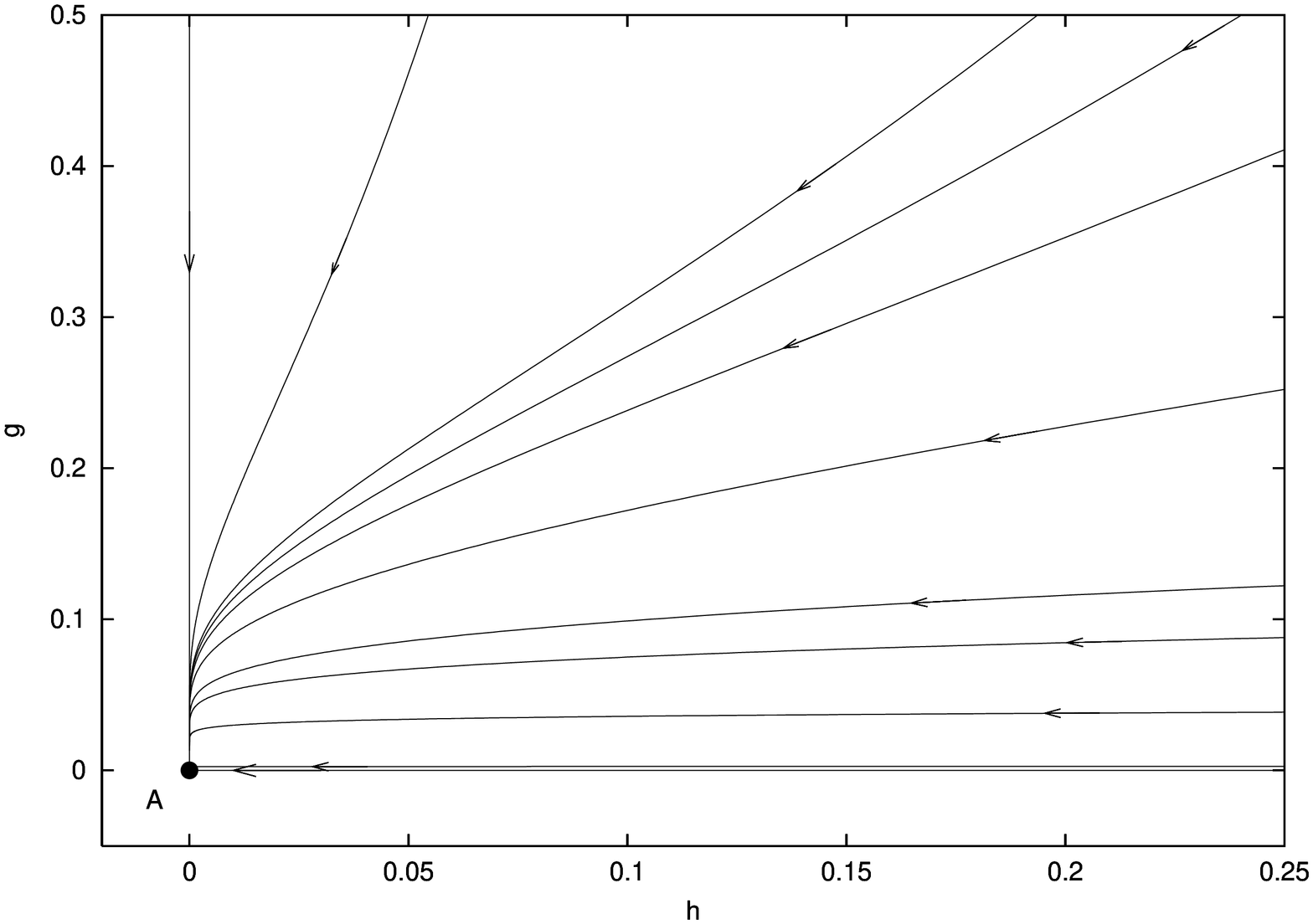}
\caption{\label{flow2} A projection of the RG flow in Eqs.
(\ref{newpairg},\ref{newpairh}). One fixed point: \textbf{A} is
attractive. This is plotted for $\epsilon = 0$.}
\end{figure}

\begin{figure}
\includegraphics[scale=0.5,angle=270]{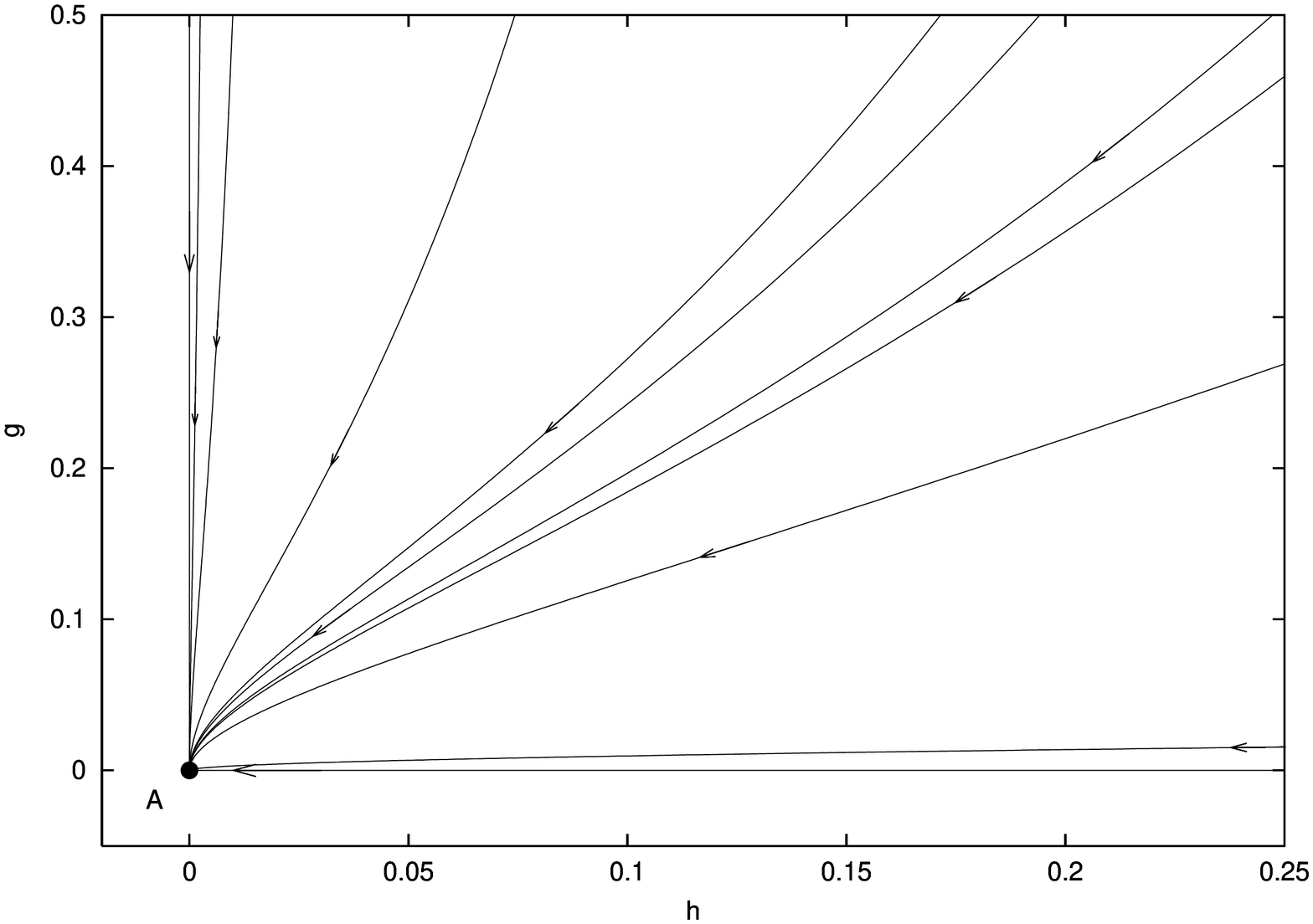}
\caption{\label{flow3} A projection of the RG flow in Eqs.
(\ref{newpairg},\ref{newpairh}). One fixed point: \textbf{A} is
attractive. This is plotted for $\epsilon < 0$}
\end{figure}

The RG flow in the regions of parameter space surrounding the
fixed points is obtained by numerically integrating the
differential equations (\ref{newpairg},\ref{newpairh}) for
various choices of distinct initial conditions chosen from within
the basins of attraction or repulsion of the fixed points. The
results are shown in the
Figs.(\ref{flow1},\ref{flow2},\ref{flow3}). The topology, the
direction of the flow and the fixed point stability are controlled
by the single parameter epsilon $\epsilon = 2 - d + y_v$. In
deriving cases (a) and (b) above, we tacitly assumed that
$\epsilon > 0$. The corresponding flow is plotted in
Fig.(\ref{flow1}). The origin is a saddle point and there is one
asymptotically stable fixed point, as shown there. For very small
initial $g$, the effective dynamics will flow towards \textbf{S}
and then be repelled to \textbf{A}. Thus, the system exhibits
crossover. In the vicinity of \textbf{S}, the correlations of the
concentrations therefore scale as ($r = |\mathbf{x}|$)
\begin{equation}\label{trivpnt}
\langle V(\mathbf{x},t)\,V(0,0)\rangle \sim r^{(y_v - d)},
\end{equation}
whereas on approaching the point \textbf{A}, they scale according
to
\begin{eqnarray}
\langle V(\mathbf{x},t)\,V(0,0)\rangle &\sim& r^{(y_v - d + 2)},
\,\, {\rm for} \,\, r \to \infty, \\
&\sim& t^{-5}, \,\, {\rm for} \,\, t \to \infty.
\end{eqnarray}

If $\epsilon = 0$, there is no fixed point at nonzero $g$. The
origin $(0,0)$ changes from a saddle point to an attracting sink
as $\epsilon \to 0$. The critical dimension is defined by
$\epsilon = 0$, and is given by $d_c = 2 + y_v$. This is the
dimension below which fluctuations are relevant. For example, for
white noise, $d_c =2$, which is also the critical dimension of the
Edwards-Wilkinson model \cite{Barabasi}.

For $\epsilon < 0$, there is only the trivial fixed point
\textbf{A} at the origin. Both these cases are shown in
Fig.(\ref{flow2}) and Fig.(\ref{flow3}), respectively. In the
neighborhood of this point, the correlations again scale as in
(\ref{trivpnt}).

%

\section{\label{sec:new3} Induced three-body reactions}

Consider the graphical representation of the stochastic equations
of motion written in (\ref{Formal_solution}) and which are
displayed below Eq.(\ref{effpropU}). In arriving at the
perturbative expansion for the vertex representing the chemical
reaction $UV^2$ in (\ref{EOM}), the steps used in the graphical
method involved replacing two of the vertex legs with the exact
expression for $V(\mathbf{k},\omega)$ and the third remaining leg
with the exact expression for $U(\mathbf{k},\omega)$ using
Eq.(\ref{Formal_solution}). This step must be iterated once more
in order to obtain all one-loop diagrams. At this intermediate
stage however we generate a set of tree diagrams (i.e., containing
no loops) with all legs terminating either in an $\eta_v$ or in an
$\eta_u$ noise factor. Now, the effective $UV^2$ vertex results
after amputating (i.e., removing) two $\eta_v$'s and one $\eta_u$
noise factor from each tree diagram and then averaging over the
remaining noise factors. The averaging step makes use of the noise
correlation functions specified in (\ref{noisecorr}). The
combinatorial factor for each diagram is found by counting all
possible noise contractions that give rise to it. In this way, we
arrived at the one-loop expansion for $\lambda$ depicted
diagramatically above in Section \ref{sec:rg}.

However, at the step leading to the set of tree diagrams, one can
instead amputate two $\eta_u$'s and one $\eta_v$ noise from the
tree diagrams and then average over the remaining noises or,
alternatively, amputate three $\eta_v$ factors and then average.
The sequence of noise factor amputations can lead to different
allowed reaction kinetics. Thus, at one-loop level, in addition to
the Gray-Scott $UV^2$ reaction, one also encounters the following
(induced) effective reactions involving a triple product of
chemical concentrations:

\begin{equation}\label{new3body}
\fbox{$\displaystyle U^2V \qquad {\rm or} \qquad V^3$ }
\end{equation}
It is important to realize that these are alternative but
\emph{exclusive} reactions, in that noise-factor amputation can
produce one reaction, say $(U^2V)$, or the other $(V^3)$, but not
both simultaneously. This is simply because each alternative
reaction derives from the same tree graph, the only difference is
due to the sequence of noise amputations and subsequent averaging
steps. Note it is not possible to amputate three $\eta_u$ factors
at one loop level (this would have led to a $U^3$ reaction, which
is therefore ruled out from the effective induced dynamics at this
order). Both reactions in (\ref{new3body}) are proportional to
$\lambda^2$ at one-loop order. Moreover, since at one loop, both
concentration fields $U$ and $V$ scale with the same exponent
$\chi$, these effective three-body reactions scale as $s^{3\chi}$
and will be relevant (or, irrelevant) if and only if the original
GS reaction is. Most importantly, the new reactions do not destroy
the scale invariance of (\ref{EOM}) at the fixed points.

Physically, this means that at one-loop order in the fluctuations,
corrections are not only induced in the original GS reaction,
leading to its renormalization (i.e., last equation in
(\ref{diffRG})), but that a new effective reaction is generated.
This new reaction is not present in the GS model (i.e., does not
exist at \textit{tree level}) but is a direct consequence of the
combined effect of noise and non-linear terms in (\ref{EOM}). In
this sense, the stochastic version of the GS is not renormalizable
at long wavelengths, because new relevant reaction terms are
thereby induced. The large scale chemistry that emerges at the
attractive fixed point $\textbf{A}$ is GS with corresponding
effective (renormalized) parameters \textit{plus} one of the
alternative induced reactions written in (\ref{newreaction1}) or
in (\ref{newreaction2}). This is interesting because it indicates
that at large scales, the coarse-grained theory (\ref{EOM})
represents an \textit{apparently} distinct chemistry.

In terms of chemical reactions, the new reaction vertices would
correspond to production of $V$ and or $U$ molecules in one of the
alternative pathways indicated below. In the case of the vertex
$U^2V$, we have various pathways
\begin{equation}\label{newreaction1}
V + 2  U \to \left\{ \begin{array}{c} 3U \\ 2V + U\\  3V
\end{array}
\right. ,
\end{equation}
whereas for the case of the vertex $V^3$,
\begin{equation}\label{newreaction2}
3V \to \left\{ \begin{array}{c} 3U \\ 2V + U\\  V + 2U
\end{array}
\right. .
\end{equation}

Thus, for the case of the the induced $U^2V$ vertex, and for the
specific pathway $V + 2U \to 2V + U$, the effective one-loop
reaction dynamics at the attractive fixed point is given by the
following deterministic equations:
\begin{eqnarray}\label{effectiveEOM1}
\frac{\partial}{\partial t} V &=& \lambda^* U V^2 +
(\lambda^*)^2U^2V + D_v \nabla^2
 V
\nonumber\\
\frac{\partial}{\partial t} U &=&  - \lambda^* U V^2
-(\lambda^*)^2U^2V- \nu^* U + D_u
 \nabla^2 U,
\end{eqnarray}
where $\lambda^*$ and $\nu^*$ denote the fixed-point values of
$\lambda$ and $\nu$, respectively. Recall that only these two
parameters renormalize nontrivially at one-loop order. Moreover,
from the fixed point analysis Case(b) given in Sec (\ref{sec:fps})
we recall the solution \textbf{A} is obtained for $\mu = 0, u_0 =
0$ and both the noise amplitudes decay to zero upon approaching
the fixed point provided $d < 12 + y$. For an attracting fixed
point, $\epsilon = 2 - d + y > 0$ which requires $d < 2 + y$,
which is the stronger inequality. Alternatively, for the case of
the one-loop induced three-body reaction $V^3$, and for the
pathway $3V \to 2V + U$, the effective one-loop reaction dynamics
at the attractive fixed point is given by the pair of
deterministic equations:
\begin{eqnarray}\label{effectiveEOM2}
\frac{\partial}{\partial t} V &=& \lambda^* U V^2 -
(\lambda^*)^2V^3  + D_v \nabla^2
 V
\nonumber\\
\frac{\partial}{\partial t} U &=&  - \lambda^* U V^2
+(\lambda^*)^2V^3- \nu^* U + D_u
 \nabla^2 U.
\end{eqnarray}

The effective reaction dynamics derived above results from the
lowest order one-loop perturbative RG calculations. If the RG
program is carried out to higher order (for example, to two loop
order), then these one-loop induced terms would have to be taken
into account.

\section{\label{sec:tempcor}Temporally correlated noise}

Starting from a microscopic description, the elimination of the
fast degrees of freedom leading to (\ref{EOM}) can in principle
result in noise with long-range correlations in both space
\textit{and} time. It is therefore of interest to investigate the
influence of temporally correlated noise on the present
phenomenological model. It is straightforward to extend the RG
analysis to incorporate long range temporal correlations. Since
the basic calculational steps are similar to those employed above,
we will be concise and list only the salient features specific to
temporal correlations.

So, in place of (\ref{noisecorr}), we consider Gaussian noise with
correlations that behave asymptotically as follows:
\begin{eqnarray}\label{tempnoise}
\left\langle \eta_v(\mathbf{k},\omega) \eta_v(\mathbf{k'},\omega')
 \right\rangle &=&
2 (2 \pi)^{d+1} A_v k^{-y_v}\omega^{-2\theta_v}
\delta^d(\mathbf{k}+\mathbf{k'}) \delta(\omega+\omega'),\nonumber
\\
\left\langle \eta_u(\mathbf{k},\omega) \eta_u(\mathbf{k'},\omega')
 \right\rangle &=&
2 (2 \pi)^{d+1} A_v k^{-y_u}\omega^{-2\theta_u}
\delta^d(\mathbf{k}+\mathbf{k'}) \delta(\omega+\omega'),
\end{eqnarray}
where the exponents $\theta_v,\theta_u$ control the range of the
temporal correlations; the limit of purely spatial correlations is
recovered by setting $\theta_v = \theta_u = 0$.

Scaling properties of the noise amplitudes are modified as
follows:
\begin{eqnarray}\label{naivetemp}
A_v &\to& s^{y_v-d-2\chi + z(1 + 2\theta_v)} A_v \nonumber \\
A_u &\to& s^{y_u-d-2\chi + z(1 + 2\theta_u)} A_u ,\nonumber \\
\end{eqnarray}
while the remainder of the relations in (\ref{naive}) are
unchanged. The steps needed to carry out the perturbation
expansion and associated diagrammatic development of (\ref{EOM})
are the same as before; only now, the noise factors appearing
there and in the diagrams are those corresponding to
(\ref{tempnoise}). The calculation of the required loop diagrams
and integrals follows the same basic steps as outlined in
Appendices A and B. Due to (\ref{tempnoise}) the integration over
internal loop frequency $ \Omega$ is much more complicated, though
still analytically tractable \cite{G&R}. The two non-trivial
one-loop differential RG equations in (\ref{diffRG}) are modified
accordingly:
\begin{eqnarray}
\frac{d\nu}{dl} &=& z \nu +
        \csc[(1 + 2\theta_v)\frac{\pi}{2}]\, \frac{\lambda A_v K_d
 \Lambda^{d-y_v}}{(\mu+D_v\Lambda^2)^{1 + 2\theta_v}}
        \nonumber \\
\frac{d\lambda}{dl}  &=&
        \left[
        2 \chi + z -
        4 \csc[(1 - 2\theta_v)\frac{\pi}{2}]\, \lambda A_v K_d \Lambda^{d-y_v}
        (\nu + D_u \Lambda^2)
        \frac{(\nu + D_u \Lambda^2)^{-1-2\theta_v} - (\mu + D_v
        \Lambda^2)^{-1-2\theta_v}}{(\mu + D_v
        \Lambda^2)^2 - (\nu + D_u \Lambda^2)^2}
\right] \lambda ,
\end{eqnarray}
and the allowed range of the noise exponent is $-\frac{1}{2} <
\theta_v < \frac{1}{2}$.
As in the case of pure spatially correlated noise, we can analyze
the RG flow and fixed points in terms of a convenient choice of
dimensionless parameters (taking $\mu = 0$ from the outset):
\begin{eqnarray}\label{dimensionless2}
g &=& \lambda A_v K_d \csc[(1 + 2\theta_v)\frac{\pi}{2}]\,
\frac{\Lambda^{d - y_v}}{\nu (D_v \Lambda^2)^{1+2\theta_v} },\\
h_1 &=& D_v \frac{\Lambda^2}{\nu},\nonumber \\
 h_2 &=& D_u
\frac{\Lambda^2}{\nu}.
\end{eqnarray}
This should be contrasted with the pair used in Section
{\ref{sec:fps}: here the temporal correlations do not allow one to
combine $h_1$ and $h_2$ consistently into a single unified
variable $(h_1 + h_2) \to h$, as we did above.

In terms of these variables, the RG equations can be written as
\begin{eqnarray}\label{triplet}
{\dot g} &=& g\left( y_v - d + 2(1+2\theta_v) - g - 4g
(1+h_2)\frac{ (\frac{h_1}{1 + h_2})^{1+2\theta_v} -
1}{(1+h_1+h_2)(-1-h_2+h_1)}
\right) \nonumber \\
{\dot h_1} &=& -h_1(2 + g)\nonumber \\
{\dot h_2} &=& -h_2(2 + g).
\end{eqnarray}
The fixed points are solved for by looking for all the zeroes of
the triplet of equations in (\ref{triplet}). As before, there are
two non-trivial fixed points and one trivial fixed point. One of
the non-trivial fixed points corresponds to $g^* = -2$, $h_1^*$
and $h_2^*$ are arbitrary constants, and we discard it for the
reasons given earlier (see comments concerning case(a) in Section
\ref{sec:fps}). The other non-trivial point corresponds to $g^* =
(y_v - d +2(1+2\theta_v))/5$ and $h_1^* = h_2^* = 0$. The
associated exponents are found to be $z^* =
-(y_v-d+2(1+2\theta_v))/5$ and $\chi^* =
(y_v-d+2(1+2\theta_v))/2$. Previous results pertaining to pure
spatially correlated noise are immediately recovered by setting
$\theta_v = 0$ in these expressions. The trivial fixed point
corresponds to $g^* = h_1^* = h_2^* = 0$, with exponents $z^* = 0$
and $\chi^* = (y_v - d)/2$, results which are seen to be identical
to those of the purely spatially correlated case. In this case,
the flow is governed by the parameter $\epsilon = y_v - d + 2(1 +
2\theta_v)$. The RG flow is qualitatively the same as before, so
in this respect, temporally correlated noise leads to no new
features at one-loop order.

\section{\label{sec:disc} Conclusions and Discussion}

We have studied the large wavelength and long time limits of the
Gray-Scott model subject to random fluctuations. We carried this
out for additive noise containing long range correlations in space
and in time, which leads one to consider a set of coupled
stochastic partial differential equations. The noise is intended
to model in part the combination of coarse-grained external
fluctuations, environmental noise and also imprecise knowledge of
initial and boundary conditions.  The asymptotic behavior of this
system is revealed from applying the dynamical renormalization
group combined with perturbation theory. This behavior is
summarized by flows in parameter space, which indicate how the
parameters of the dynamic model change under coarse-graining;
general points in this space represent the dynamics
\emph{effective} at the corresponding space and time scale. The
fixed points control the effective dynamics at the largest scales
and corresponding correlation functions (in terms of the chemical
concentrations or composition fields) are power laws depending on
a pair of scaling exponents. It will be noted that the information
obtained here is of a \emph{statistical} nature, since here one
deals in correlation functions, which are themselves
probabilities. These correlations provide a measure of the long
wavelength properties of the patterns that can form in the
presence of fluctuations.

The RG calculation was carried out to one-loop order. At this
lowest order, only two out of eight of the model parameters
receive corrections, yet there is already a non-trivial infrared
stable fixed point and the phenomena of cross-over. We have also
identified the critical dimension below which fluctuations are
relevant. This dimension is $d_c = 2 + y_v$ for spatially
correlated noise or $d_c = 2 + y_v +2(1 + 2\theta_v)$, for noise
with both spatial and temporal correlations, respectively. In the
limit of large scales (after coarse-graining) the properties of
the system converge towards a simple solution with fractal
properties (i.e., exhibiting scale-invariance) as shown in
Fig(\ref{flow1}). The existence of this solution depends on the
sign of $\epsilon$. The non-trivial scale invariant solution is
present only for positive values of $\epsilon$, when the
fluctuations become important and then emergent behavior arises.
Moreover, we find that the combined interplay of the nonlinearity
and the fluctuations lead to effective three-body reaction terms
not present in the original GS model. At one-loop order in the
fluctuations the new reaction vertices correspond to three-body
molecular reactions with various alternative pathways as expressed
in Eqs (\ref{newreaction1}) and (\ref{newreaction2}).

The renormalized reaction-diffusion equations therefore can be
represented by a chemistry apparently distinct from the original
GS reaction. For small noise amplitudes, a one-loop calculation
should be adequate to capture some of the salient features of
stochastic reaction-diffusion dynamics. In fact, simple white
noise can lead to rather striking effects. Preliminary numerical
calculations demonstrate clearly that both pattern
\emph{selection} and dynamic pattern \emph{replication} can be
controlled by adding white noise to the deterministic GS model
\cite{LHMP2}. These studies raise important questions pertaining
to the role of noise in both chemical and biological
self-organization and the environmental selection of emergent
properties.

At one-loop order, temporally correlated noise leads to no
noteworthy features in the RG analysis, though that conclusion may
change at next higher order. Temporally correlated fluctuations
can lead to complex scaling exponents which signal the presence of
hierarchical structures \cite{JPM}. Such structures are absent at
one loop order.

\begin{acknowledgments}
We thank Marcel Vlad for reading a preliminary version of this
paper and for providing us with useful comments. The research of
FM is supported in part by the Grant \# BMC2000-0764 from MCyT
(Spain). FL acknowledges a fellowship provided by INTA for
training in astrobiology.  D.H. and J.P-M. are supported by funds
from INTA, Comunidad Aut\'onoma de Madrid and Grant \#
BXX2000-1385 from MCyT (Spain).
\end{acknowledgments}

\appendix
\section{Propagator Renormalization}

Details of the propagator renormalization are give here. As the
diagrams for the expanded propagators show, at one loop only the
$U$ propagator receives corrections. Reverting back from diagrams
to corresponding algebraic quantities, we have that
\begin{equation}\label{Uprop}
G_u(\mathbf{k},\omega) = G_{u0}(\mathbf{k},\omega) + \lambda \,
[G_{u0}(\mathbf{k},\omega)]^2 \int_{\mathbf{p},\Omega} 2 A_v
p^{-y_v}G_{v0}(\mathbf{p},\Omega)\, G_{v0}(-\mathbf{p},-\Omega),
\end{equation}
where $\int_{\mathbf{p},\Omega} = \int \frac{d^d
\mathbf{p}}{(2\pi)^d} \int_{-\infty}^{\infty}
\frac{d\Omega}{2\pi}$ is an abbreviation for the integration over
wave-vector and frequency.  The integration over wave-vector can
of course be further decomposed into an integration over angles
and modulus. Eq(\ref{Uprop}) can be used to obtain an expansion
for the \emph{inverse} propagator:
\begin{equation}\label{invUprop}
G_u^{-1}(\mathbf{k},\omega) = G_{u0}^{-1}(\mathbf{k},\omega) -
\lambda \,\int_{\mathbf{p},\Omega} 2 A_v
p^{-y_v}G_{v0}(\mathbf{p},\Omega)\, G_{v0}(-\mathbf{p},-\Omega) +
O(\lambda^2).
\end{equation}
Referring back to the structure of the bare propagator in
(\ref{Bare_propagator}), in order that the original model
(\ref{EOM}) be renormalizable in the hydrodynamic limit, the
inverse effective propagator must have the form
\begin{equation}\label{effective-Uprop}
G_u^{-1}(\mathbf{k},\omega) = \tilde{\nu} + \tilde{D_u} k^2 -
i\omega,
\end{equation}
where $\tilde{\nu}$ and $\tilde{D_u}$ are the effective, or
renormalized decay rate and diffusion constants for the $U$ field,
respectively.

Since the one-loop correction integral in (\ref{invUprop}) does
not depend on either $\omega$ nor $k$, we can immediately conclude
that the diffusion constant is unrenormalized at one loop order,
$\tilde{D}_u = D_u$ while the decay rate renormalization is given
by
\begin{equation}\label{nuren}
\tilde{\nu} = \nu + \lambda \,\int_{\mathbf{p},\Omega} 2 A_v
p^{-y_v}G_{v0}(\mathbf{p},\Omega)\, G_{v0}(-\mathbf{p},-\Omega).
\end{equation}
The frequency integration over $\Omega$ can be performed exactly
(e.g., by residues). Finally, eliminating a finite band of large
wave-numbers below the cutoff $\Lambda$ in the Wilsonian fashion
then yields
\begin{equation}
\tilde{\nu}^{<} = \nu + \lambda\, A_{v} \,K_d
\int^{\Lambda}_{\Lambda/s} dp \, \frac{p^{d-y_v -1}}{(\mu +
D_{v}p^2)}.
\end{equation}
For an infinitesimally thin wave-number shell ($s = 1 + \delta$,
$0 < \delta << 1$ ) we pass from an integral to a differential
relation. After a further re-scaling according to (\ref{naive}) we
obtain the first differential RG equation as displayed in
(\ref{diffRG}).

\section{Vertex Renormalization}

Steps similar to those above are involved in the renormalization
of the one-loop vertex (or, coupling $\lambda$). Transcribing the
vertex diagram in the text back into algebraic quantities, we have
that
\begin{equation}
\tilde{\lambda} = \lambda - 4\lambda^2 \,\int_{\mathbf{p},\Omega}
2A_{v} p^{-y_v} G_{v0}(\mathbf{p},\Omega) \,
G_{v0}(-\mathbf{p},-\Omega)\, G_{u0}(-\mathbf{p},-\Omega),
\end{equation}
where we have set all external wave-numbers and frequencies to
zero from the outset, in anticipation of the hydrodynamic limit.
As before, the frequency integration can be performed immediately
by means of residues. Eliminating a finite band of large
wave-numbers yields
\begin{equation}
\tilde{\lambda}^{<} = \lambda - 4\lambda^2 \, A_{v}\, K_d \,
\int_{\Lambda/s}^{\Lambda} dp \, \frac{p^{d-y_v -1}}{(\mu +
D_{v}p^2)(\mu + \nu + [D_{u} + D_{v}]p^2)},
\end{equation}
For an infinitesimally thin wave-number shell ($s = 1 + \delta$,
$0 < \delta << 1$ ) we pass from an integral to a differential
relation. After a further re-scaling according to (\ref{naive}) we
obtain the last differential RG equation as displayed in
(\ref{diffRG}).

\end{document}